\documentclass[12pt]{JHEP3}
\usepackage{epsfig,amsmath}

\def\a               {\alpha}
\def\b               {\beta}
\def\d               {\delta}

\def\t               {\theta}

\def\x               {\chi}
\def\D               {\Delta}

\def\ISAJET          {{\small ISAJET\,7.64}}

\def\SOFTSUSY        {{\small SOFTSUSY\,1.71}}
\def\SPHENO          {{\small SPHENO\,2.0}}
\def\SUSPECT         {{\small SUSPECT\,2.101}}

\def\ISAJETnn        {{\small ISAJET}}
\def\ISASUSYnn       {{\small ISAJET}}
\def\SOFTSUSYnn      {{\small SOFTSUSY}}

\def\SUSPECTnn       {{\small SUSPECT}}

\def\ti              {\tilde}

\def\sq              {\ti q}
\def\st              {\ti t}
\def\sb              {\ti b}
\def\ch              {\ti \x^\pm}

\def\nt              {\ti \x^0}
\def\sg              {\ti g}

\newcommand{\mst}[1]   {m_{\st_{#1}}}

\newcommand{\mch}[1]   {m_{\ti \x^+_{#1}}}
\newcommand{\mnt}[1]   {m_{\ti \x^0_{#1}}}

\newcommand{\msg}      {m_{\ti g}}

\def\gev             {{\rm GeV}}

\def\DR              {{\rm\overline{DR}}}
\def\MS              {{\rm\overline{MS}}}

\def\noi {\noindent}
\def\nn  {\nonumber}

\newcommand{\eq}[1]   {\mbox{(\ref{eq:#1})}}
\newcommand{\fig}[1]  {\mbox{fig.~\ref{fig:#1}}}
\newcommand{\Fig}[1]  {\mbox{Figure~\ref{fig:#1}}}
\newcommand{\sect}[1] {\mbox{section~\ref{sec:#1}}}

\newcommand{\gsim}{\;\raisebox{-0.9ex}
           {$\textstyle\stackrel{\textstyle >}{\sim}$}\;}
\newcommand{\lsim}{\;\raisebox{-0.9ex}{$\textstyle\stackrel{\textstyle<}
           {\sim}$}\;}

\title{Theoretical uncertainties in sparticle mass predictions from
  computational tools}

\author{B.C. Allanach$^{1}$, S. Kraml$^2$ and W. Porod$^3$\\
$^{1}$ LAPTH, Chemin de Bellevue, B.P. 110, F-74941 Annecy-le-vieux, France\\
$^{2}$ CERN Theory Division, CH-1211 Geneva 23, Switzerland\\
$^3$ Physik Inst., Univ. Zurich, Winterthurerstrasse 190, CH-8057 Zurich, Switzerland}

\abstract{
  We estimate the current theoretical uncertainty in sparticle mass 
predictions by comparing several state-of-the-art 
computations within the minimal supersymmetric standard model (MSSM).
We find that the theoretical uncertainty is comparable to the expected
statistical errors from the Large Hadron Collider (LHC), 
and significantly larger than those 
expected from a future $e^+e^-$ Linear Collider (LC). 
We quantify the theoretical uncertainty on relevant sparticle
observables for both LHC and LC, and show that the value of the error is
significantly dependent upon the supersymmetry (SUSY) breaking
parameters. 
We also present the theoretical uncertainty induced in fundamental
scale SUSY breaking parameters when they are fitted from LHC measurements.
Two regions of the SUSY parameter space where accurate predictions 
are particularly difficult are examined in detail: the large $\tan \beta$ 
and focus point regimes.}

\preprint{CERN-TH/2003-029\\ LAP-TH-963/03\\ ZU-TH 03/03\\ hep-ph/0302102}

\begin{document}
\bibliographystyle{JHEP}

\section{Introduction}
Weak-scale supersymmetry (SUSY) is well
motivated~\cite{Haber:1993wf,Martin:1997ns} because 
it solves the 
technical hierarchy problem, removing ultra-violet quadratic divergent
corrections to the Higgs mass. 
Viable weak-scale supersymmetry
implies over a hundred ``soft breaking'' terms
in the MSSM Lagrangian at the weak scale, making general analysis intractable.
By making model-dependent assumptions on the origin of SUSY
breaking for the MSSM fields, one often provides relations between
the weak-scale SUSY breaking parameters, vastly reducing the available
parameter space. Many theoretical schemes of SUSY breaking exist, the most
popular of which can be classified by the mechanism that mediates SUSY
breaking from a hidden sector to the MSSM fields. 
In this paper, we will use 
minimal supergravity (mSUGRA)~\cite{Abel:2000vs}, minimal anomaly mediation
(mAMSB)~\cite{Gherghetta:1999sw} and minimal 
gauge (mGMSB)~\cite{Kolda:1998wt} mediation.

Supposing SUSY is discovered at a present or future collider,
it will be a major challenge to measure the SUSY breaking parameters with 
good accuracy. 
In order to determine the free parameters, physical sparticle masses (or
kinematical variables related to them) will be measured. These sparticle
masses must be turned into SUSY breaking parameters, 
typically at some high scale.
Provided enough information is collected, this will allow
tests on the relations between the high-scale parameters.
For example, a simple question already
addressed~\cite{Tsukamoto:1995gt,Blair:2000gy,Blair:2002pg,Zerwas:2002as} 
is: do the gaugino masses unify, and if so, do they unify at the same scale as
the gauge couplings? A double affirmative would be strong support in favour of 
SUSY GUT-type schemes. On the other hand, there are many other possibilities:
for example, string models can be non-universal~\cite{Brignole:1994dj}, 
or unify at an intermediate scale. 

There is a large literature upon expected empirical errors on the
observables both a Linear
Collider~\cite{Kuhlman:1996rc,Baer:1996vd,Freitas:2002gh,Aguilar-Saavedra:2001rg,Abe:2001gc} 
and the LHC~\cite{ATLASTDR,Denegri:1999pe,lesHouches,Branson:2001ak}.
It will be essential to know the
theoretical uncertainty if we are to discriminate models of SUSY breaking.
There are several studies of how empirical errors propagate
into uncertainties on fundamental-scale SUSY breaking parameters in the 
literature~\cite{Hinchliffe:1997iu,Hinchliffe:1998ys,Bachacou:1999zb,
ATLASTDR,Blair:2000gy,Allanach:2001xz,Blair:2002pg,Zerwas:2002as}. 

The deduction of high-scale SUSY breaking parameters from
observables inevitably involves theoretical errors coming from the
level of approximation used (e.g.\ neglected higher order terms), and it is
these uncertainties that we study here. 
We use four modern codes to calculate MSSM spectra:
\ISAJET~\cite{Baer:1999sp}, \SOFTSUSY~\cite{Allanach:2001kg},
\SPHENO~\cite{Porod:2003um} and \SUSPECT~\cite{Djouadi:2002ze}. 
We use the differences in results between the codes to define the 
current theoretical uncertainties.
We have done our best to eliminate differences due to bugs by examining 
the relevant parts of codes in detail if there was an obvious large 
discrepancy. However, it would be unrealistic to
claim that all of the codes are completely bug free. We therefore take a 
practical interpretation of `theoretical uncertainty': after all, 
when fitting experimental data to SUSY breaking models, one must use one of
the available computational tools. 
The precise implementation of the known higher-order corrections 
differs and has been found to produce significantly different results (for
example, using different scales for parameters in the highest-order corrections).
Therefore, certainly the differences in results
between the codes are due (at least to a large part) to unknown higher-order
corrections, matching more traditional notions of `theoretical uncertainty'. 

Previously, there have been some studies of differences between various 
calculations of the MSSM spectrum. 
2-5$\%$ differences in sparticle masses were noticed~\cite{me1} between 
various particles along the Snowmass (SPS)~\cite{Allanach:2002nj} 
model lines~1a and 6 between the \SOFTSUSYnn\,1.2, \ISAJETnn\,7.51~and
\SUSPECTnn\,2.0~programs. Model line~2 (the so-called focus point line, 
with very heavy scalar sparticles) was observed to show huge 30\% 
differences in the masses of the weak gauginos. 
Differences in the spectra and branching ratios were also observed 
in~\cite{Ghodbane:2002kg} comparing {\ISAJETnn\,7.58}, 
{\small SUSYGEN\,3.00} and {\small PYTHIA\,6.2}. 
Moreover, large 10$\%$ level differences between \SOFTSUSYnn\,1.3, 
\ISAJETnn\,7.58 and {\small feynSSG}~were observed~\cite{lesHouches} 
in the mSUGRA Post-LEP benchmarks~\cite{Battaglia:2001jn}. 
Some of the benchmarks in some of the codes were
found to not be consistent with electroweak symmetry breaking, unless one
fiddled with the Standard Model inputs $m_t$ and $\alpha_s(M_Z)$. These were
noticeably points E,F (focus-points) and K,M (high $\tan \beta$ and high
$m_0, m_{1/2}$ points). 
Likewise, the input parameters had to be adjusted in \cite{Battaglia:2001jn} 
to get similar spectra from \ISASUSYnn\,7.51 and {\small SSARD}.
Initial results highlighting the differences between the predictions in the
focus-point and high $\tan \beta$ regimes have already been presented by the
current authors as a conference proceeding~\cite{susy02}.
We will  present the main results of this last work here for completeness,
updated to state-of-the art calculations.

In this paper, we push these initial observations of theoretical uncertainties
further by (i) comparing them to the expected experimental accuracies at the
LHC and a future LC  and (ii) examining their 
effect upon future empirical fits to fundamental-scale SUSY breaking schemes. 
We take expected statistical errors upon sparticle observables from previous
studies of mSUGRA at the LHC~\cite{ATLASTDR} at two of the LHC
benchmark points.
We then perform a fit to mSUGRA at each point with each of the four codes.
The statistical precision of the fitted fundamental mSUGRA parameters 
may then be compared with the theoretical error by looking at the 
differences between the four fits. 
We next quantify and present the theoretical error in the coloured
sparticle masses at the SPS points. 
We also re-examine the SPS points providing theoretical uncertainties 
on mass predictions for sparticles that may kinematically be accessed 
at a 500 GeV LC.
Quantification of errors at particular points will give
us an idea of their magnitude and whether or not they significantly depend
upon the SUSY breaking scheme (or point). 

In \sect{codes}, we introduce the four state-of-the art
calculations we will use and their level of approximation: \ISAJET,
\SOFTSUSY, \SPHENO\ and \SUSPECT.  
`Tricky' regions of the MSSM parameter space, where it is difficult 
to make accurate predictions, are discussed in \sect{tricky}: 
large $\tan\beta$ and the focus-point regime.
In \sect{lhc}, we perform the LHC mSUGRA empirical fits and then quantify 
the theoretical error on squark and gluino masses for the SPS points.  
In \sect{lc}, we quantify theoretical errors on masses relevant for a 
500~GeV Linear Collider. Finally, there are conclusions 
and an outlook in \sect{conc}.

\section{The codes \label{sec:codes}}

We compare the latest versions of four public SUSY renormalisation group
evolution (RGE) codes 
which we think constitute a representative sample of such programs: 
\ISAJET, \SOFTSUSY, \SPHENO\ and \SUSPECT.
The basic principle of the SUSY mass spectrum 
calculation is the same in all programs:
Gauge and Yukawa couplings are taken as input parameters at the 
electroweak scale. However, in the MSSM, in order to define them in the
$\DR$ scheme, from experimental observables, one must first subtract
threshold corrections from sparticles. The sparticle spectrum is unknown at
this stage and so to begin the calculation, some guess is made for the soft
SUSY breaking parameters and spectrum.
The MSSM parameters are then run to the high scale $M_X$ by RGEs. 
At $M_X$, boundary conditions are imposed on the SUSY breaking parameters.
Couplings and SUSY parameters are then run back down to 
$M_{SUSY} \equiv \sqrt{m_{{\tilde t}_1} m_{{\tilde t}_2}}$ where the $\mu$ and
$B$ MSSM Higgs potential parameters are set in order to give correct radiative
electroweak symmetry 
breaking consistent with an input value of $\tan \beta=v_2/v_1$ 
($v_1$, $v_2$ being the two Higgs fields' vacuum expectation values (VEVs)).
The SUSY masses are calculated and radiative corrections are 
applied, and the parameters are run down to the electroweak scale. Finally, 
the whole process is iterated in order to obtain a stable solution.

An overview of which corrections are implemented in each of 
the four programs is given in Table~\ref{tab:codes}.
From this table we already expect some differences in the results 
due to the different levels of radiative corrections applied. 
In particular \ISAJET\ has no finite radiative corrections 
to most of the sparticle masses. 
However, we note that even if each column of the table were identical, one
could expect different numerical results from each of the codes. 
In practise, if a quantity is calculated at one-loop, one has the freedom 
(at one-loop accuracy) of using whatever scale one desires for parameters 
in the one-loop correction itself. 
The difference between using, for instance, pole or running masses 
or couplings derived in the $\MS$ or $\DR$ scheme in one-loop corrections 
is formally of higher-loop order, but leads to non-negligibly different 
results. 
Indeed, one can roughly estimate the effects of
higher-loop terms by the difference in predicted masses by varying 
the scale of parameters in the highest loop included.

\begin{footnotesize}
\TABULAR[t]{c|c|c|c|c}{
    & \ISAJET & \SUSPECT & \SOFTSUSY & \SPHENO  \\
  \hline
  \bf RGEs & 2--loop & 2--loop & 2--loop & 2--loop  \\
           &         & scalars at 1--loop & & \\
  \hline
  \bf VEVs & not running & \multicolumn{3}{c}{running (1--loop)}  \\
  \hline
  \bf Yukawa cpl. & & & \\
  $h_t$ & full 1--loop & full 1--loop & full 1--loop & full 1--loop \\
  $h_b$ & full 1--loop          
        & \multicolumn{2}{c}{$bg + \sb\sg + \st\ch$ loops} 
        & full 1--loop \\
  \hline
  \bf Higgs sector & & \multicolumn{2}{c}{} \\
    tadpoles & 3rd gen.\ (s)fermions 
      & \multicolumn{2}{c|}{complete 1--loop \cite{Pierce:1997zz}} 
      & 2-loop \cite{Pierce:1997zz,Dedes:2002dy} \\
    $h^0$, $H^0$ & 1--loop  \cite{Bisset:1995dc} 
                 & var. ops. 
                 & 2--loop  \cite{Heinemeyer:1998yj}
                 & 2--loop  \cite{Degrassi:2001yf,Brignole:2001jy} \\
  \hline
  \bf SUSY\ masses & & & \\ 
    $\ch,\nt$ & some corr. for $\ch_1$ 
      & \multicolumn{2}{c|}{1--loop aprox. for $\D M_1$, $\D M_2$, $\D\mu$} 
      & full 1--loop \\
    $\st$ & --- & 1--loop approx. & full 1--loop & full 1--loop \\
    $\sb$ & --- & 1--loop approx. & full 1--loop & full 1--loop \\ 
    $\sg$ & \multicolumn{4}{c}{all have 1-loop $g\sg + q\sq$ re-summed} \\
  \hline}
{\label{tab:codes}RGEs and radiative corrections implemented in \ISAJET, 
  \SUSPECT, \SOFTSUSY~and \SPHENO.}
\end{footnotesize}

Let us now point out some of the differences in more detail. 
A subtle but important issue is the treatment of Yukawa couplings. 
In general, they are derived from the quark masses as
\begin{equation}
   h_t = \sqrt{2}\,m_t/v_2,\qquad h_b = \sqrt{2}\,m_b/v_1,
\end{equation}
where $m_t$ and $m_b$ are the running top 
and bottom quark masses in the $\DR$ scheme. 
Obviously, it makes a difference whether the VEVs $v_{1,2}$ are 
running (\SOFTSUSY, \SPHENO, \SUSPECT) or not (\ISAJET). 
It makes  an important difference how the $\DR$ masses
are calculated from the pole or $\MS$ masses. 
We first discuss Standard Model threshold corrections, and 
afterwards the sparticle loop corrections.

\noi $\bullet$ 
\ISAJET\ takes 2-loop QCD corrections to $m_t$ into account,
including the shift from the $\MS$ to the $\DR$ scheme as~\cite{Avdeev:1997sz} 
\begin{equation}
  m_t(M_t)^\DR_{\rm SM} = M_t \left[ 1 + \frac{5\a_s}{3\pi} + 
  \left(16.11-1.04 \left(5 - \frac{6.63}{M_t}\right)\right)
  \left(\frac{\a_s}{\pi}\right)^2 \right]^{-1},
\label{eq:dmtIsa}
\end{equation}
where $M_t$ is the top pole mass and 
$\a_s=\a_s(M_t)$ at 3-loops in the $\MS$ scheme.
For the bottom quark mass, \ISAJET\ takes a hard-coded 
value of $m_b(M_Z)^\DR_{\rm SM}=2.82$~GeV~\cite{Baer:2002ek}.

\noi $\bullet$ 
\SOFTSUSY\ and \SPHENO\ calculate both $h_t$ and $h_b$ at $Q=M_Z$. 
The $\DR$ top mass is related to the pole mass by 
2-loop QCD \cite{Avdeev:1997sz,Bednyakov:2002sf}:
\begin{equation}
  m_t(Q)^\DR_{\rm SM} = M_t \,  
  \left[ 1 - \frac{\a_s}{3\pi}\left(5-3L\,\right) 
           - \a_s^2\left(0.538 - \frac{43}{24\pi^2}\,L
                               + \frac{3}{8\pi^2}\,L^2 \right)\right]
\label{eq:mtSMsoft}
\end{equation}
where%
%
\footnote{After the publication of this paper, it was noted that 
eq.~\eq{mtSMsoft} incorrectly describes the 2-loop QCD corrections 
to the top mass: eq.~\eq{mtSMsoft} holds for $L=\ln(m_t^2(Q)/Q^2)$,  
not for $L=\ln(M_t^2/Q^2)$. 
\SOFTSUSY\ and \SPHENO\ therefore contain the incorrect formula, 
which ought to be 
\begin{equation}
  m_t(Q)^\DR_{\rm SM} = M_t \,  
  \left[ 1 - \frac{\a_s}{3\pi}\left(5-3L\,\right) 
           - \a_s^2\left(0.876 - \frac{91}{24\pi^2}\,L
                               + \frac{3}{8\pi^2}\,L^2 \right)\right] 
  \tag{2.3a}
\end{equation}
for $L=\ln(M_t^2/Q^2)$. 
We thank D.\,R.\,T.\ Jones for drawing our attention to this issue.} 
%
%
$L=\ln(M_t^2/Q^2)$. 
For the $b$ quark, 3-loop relations~\cite{Melnikov:2000qh} 
are used to calculate $m_b(M_b)^\MS_{\rm SM}$ from the $b$ pole mass, 
which is run to $M_Z$ by 3-loop RGEs~\cite{Tarasov:1980au,Gorishnii:1984zi}. 
$m_b(M_Z)^\MS_{\rm SM}$ is then related to the $\DR$ mass 
by~\cite{Allanach:2001kg}   
\begin{equation}
  m_b(M_Z)^\DR_{\rm SM} = m_b(M_Z)^\MS_{\rm SM}  
  \left[ 1 - \frac{\a_s}{3\pi} - \frac{35\a_s^2}{72\pi^2} 
         + \frac{3g_2^2}{128\pi^2} + \frac{13g_1^2}{1152\pi^2}\,\right] 
\label{eq:mbSMsoft}
\end{equation}
in \SOFTSUSY. \SPHENO\ neglects the last two terms of eq.~\eq{mbSMsoft}.

\noi $\bullet$ 
\SUSPECT\ uses 2-loop relations and 2-loop RGEs  
to derive $m_b(M_b)^{\MS}_{\rm SM}$ and $m_t(M_t)^{\MS}_{\rm SM}$ 
from the $b$ and $t$ pole masses. For the conversion to the $\DR$ 
scheme, eqs.~\eq{mtSMsoft} and \eq{mbSMsoft} are applied 
(with $L=0$ in eq.~\eq{mtSMsoft} since $Q=M_t$). \\
Using $Q=M_t$ to define $m_t(Q)$ (\ISAJET, \SUSPECT) is in principle more
accurate than using $Q=M_Z$ (\SOFTSUSY, \SPHENO) since then the QCD logs
between $M_Z$ and $M_t$ are re-summed.  
We refer the reader to the respective
manuals~\cite{Baer:1999sp,Allanach:2001kg,Porod:2003um,Djouadi:2002ze} for more
details. 
 
The next step is to include SUSY loop corrections. 
For $m_t$, all four programs apply full 1-loop SUSY corrections 
according to \cite{Pierce:1997zz}. For $m_b$,  the full 1-loop 
(\ISAJET, \SPHENO) or the leading (\SOFTSUSY, \SUSPECT) SUSY 
corrections are included. The $\tan\b$ enhanced corrections to $m_b$ 
are re-summed as given in \cite{Carena:1999py} in all four programs, 
see~\cite{susy02}. 
Still, there are some important differences in the implementation 
of these corrections. For example, the $\sg\st$ correction to 
$m_t$ is
\begin{eqnarray}
  \left(\frac{\Delta m_t}{m_t}\right)^{\sg\st} 
  &=& -\frac{\a_s}{3\pi}\,\Big\{ 
      B_1(m_t^2,\msg^2,\mst{1}^2) + B_1(m_t^2,\msg^2,\mst{2}^2) \nn \\
  & & \hspace{12mm}+\,\frac{\msg}{m_t}\,\sin2\t_{\st} 
      \left[B_0(m_t^2,\msg^2,\mst{1}^2)-B_0(m_t^2,\msg^2,\mst{2}^2)\right]
      \Big\} .
  \label{eq:dmt}
\end{eqnarray}
\ISAJET\ adds this correction at $m_t$, but using a scale 
$Q=\sqrt{\mst{L}\mst{R}}$, (this is done to avoid double counting of 
logarithmic corrections which are included via step functions in the RGEs)
and $\a_s=\a_s(\msg)$ in eq.~\eq{dmt}, 
while \SOFTSUSY\ and \SPHENO\ calculate it at $Q=M_Z$ and \SUSPECT\ 
at $Q=M_t$.
Accordingly, $m_t=m_t(Q)^{\DR}_{\rm SM}$ in the term $\msg/m_t$
enters with different values in all four programs.
Due to differences in the inclusion 
of finite radiative corrections to sparticle masses, the gluino 
and stop masses in \eq{dmt} vary from program to program.
In particular, \ISAJET\ calculates the corrections to $\msg$ 
at $Q=\msg$ while \SUSPECT\ calculates them at $Q=M_Z$, which 
leads to quite different gluino masses. 
Analogous differences exist in the other contributions  
to $(\Delta m_t)^{SUSY}$ as well as in the calculation of 
$(\Delta m_b)^{SUSY}$.

Another comment is in order concerning $\a_s$.
The value of $\a_s(M_Z)$ from experiment is given in the $\MS$
scheme. \SOFTSUSY, \SPHENO\ and \SUSPECT\ take $\a_s(M_Z)^{\MS}$ 
as input and convert it to the $\DR$ scheme~\cite{Pierce:1997zz}
\begin{equation}
   \a_s^\DR(M_Z) = \frac{\a_s(M_Z)^{\MS}}{1-\D\a_s} \,,
\end{equation}
where 
\begin{equation}
   \D\a_s = \frac{\a_s(M_Z)}{2\pi} \left[ \frac{1}{2} 
      - \frac{2}{3}\ln\left(\frac{M_t}{M_Z}\right) 
      - 2\ln\left(\frac{\msg}{M_Z}\right)
      - \frac{1}{6}\sum_{\sq}\sum_{i=1,2}\ln\left(\frac{m_{\sq_i}}{M_Z}\right)
   \right]\,.
\label{eq:das}
\end{equation}
In \SUSPECT, the log terms are not added explicitly but included via 
threshold functions in the RGEs. 
Also \ISAJET\ re-sums the logs in \eq{das} via step-function decoupling 
in the RGEs. The finite term, however, is not taken into account. 
The difference due to the finite term is small but relevant (1\%).
One could in principle interpret the input $\a_s(M_Z)$ in \ISAJET\ as 
already being the effective Standard Model $\DR$ value. 
Since, however, for some corrections \ISAJET\ effectively takes hard-wired 
values of $\a_s(M_Z)^{\MS}=0.118$, we take the canonical input 
value of $\a_s(M_Z)^{\MS}=0.118$ for all codes. 


\section{Tricky corners of SUSY parameter space \label{sec:tricky}}

As a general point,
when quantifying errors on predicted observables, 
we assume that the results from the codes follow a Gaussian probability
distribution. This is {\em a priori} unjustified, but we prefer it to
quoting minimum and maximum values because we believe that the additional 
information included in the variance is desirable.
For example, if three codes all provided identical results and one gave a
result further away, we think that the true uncertainty ought to be less
than the range of minimum-maximum results.

\subsection{Large $\tan \beta$}

Large $\tan\b$ has always been recognised as a difficult case since  
it requires a thorough treatment of the bottom Yukawa coupling.  
\Fig{yb} shows $h_b$ of the four different programs 
as a function of $\tan\b$ in the mSUGRA model. 
We see that \SOFTSUSY\ and \SPHENO\ agree very well on $h_b$,   
and there is also good agreement with \ISAJET.
Comparing only these programs we would assign a $\lsim 3\%$ 
uncertainty on $h_b$ even for very large $\tan\b$.
The agreement with \SUSPECT\ is however not so good, and we find 
4\,--\,8\% uncertainty taking all four programs into account. 
The effect of re-summing the $\tan\b$ enhanced SUSY loop corrections 
can be seen when comparing the solid and dotted lines in \fig{yb}, 
the solid line being the result of \ISAJET, where the $(\Delta m_b)^{SUSY}$ 
corrections are resummed, and the dotted one being the result of 
{\small ISASUSY\,7.58}, where this re-summation is not applied. 

\FIGURE[t]{%
\unitlength=1mm
\begin{picture}(145,68)
\put(40,3){\mbox{\epsfig{figure=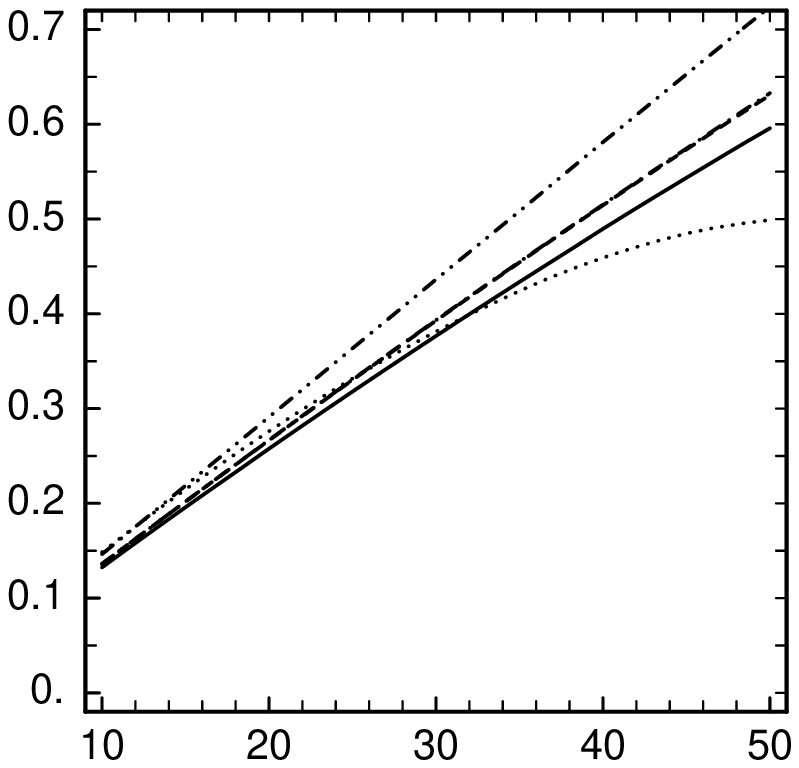,height=6.4cm}}}
\put(34,28){\rotatebox{90}{$h_b\,(M_{SUSY})$}}
\put(70,0){\mbox{$\tan\b$}}
\end{picture}
\label{fig:yb}
\caption{Bottom Yukawa coupling at $M_{SUSY}$ scale 
  as a function of $\tan\b$, for 
  $m_0=400$~GeV, $m_{1/2}=300$~GeV, $A_0=0$, $\mu>0$, $M_t=175$~GeV; 
  full (dotted) lines: \ISAJET~(7.58), dashed: \SOFTSUSY, 
  dash-dotted: \SPHENO, dash-dot-dotted: \SUSPECT.}} 

The bottom Yukawa coupling has its largest effect in the Higgs sector when it
is large (at high $\tan \beta$): 
the evolution of $m_{H_1}^2$ is driven by $h_b$,
\begin{equation}
  \frac{dm_{H_1}^2}{dt} \sim \frac{3}{8\pi^2}\,h_b\,X_b +\ldots\,,\quad  
  X_b = (m_{\ti Q}^2 + m_{\ti D}^2 + m_{H_1}^2 + A_b^2)\,, 
\end{equation}
where $t=\ln Q$, $Q$ being the renormalisation scale.
Differences in $m_{H_1}^2$ directly translate into the physical 
Higgs boson masses since
\begin{equation}
   m_A^2 = \frac{1}{c_{2\b}}
   \left( \overline{m}_{H_2}^2 - \overline{m}_{H_1}^2\right) 
   + \frac{s_\b^2\,t_1}{v_1} + \frac{c_\b^2\,t_2}{v_2} - M_Z^2  \,.
\label{eq:mA}
\end{equation}
Here $\overline{m}_{H_i}^2 = m_{H_i}^2 - t_i/v_i$, $i=1,2$, and 
$t_{1,2}$ are the tadpole contributions. The self energies of 
$Z$ and $A$ have been neglected in eq.~\eq{mA}.

\FIGURE[t]{%
\unitlength=1mm
\begin{picture}(145,135)
\put(4,0){\epsfig{file=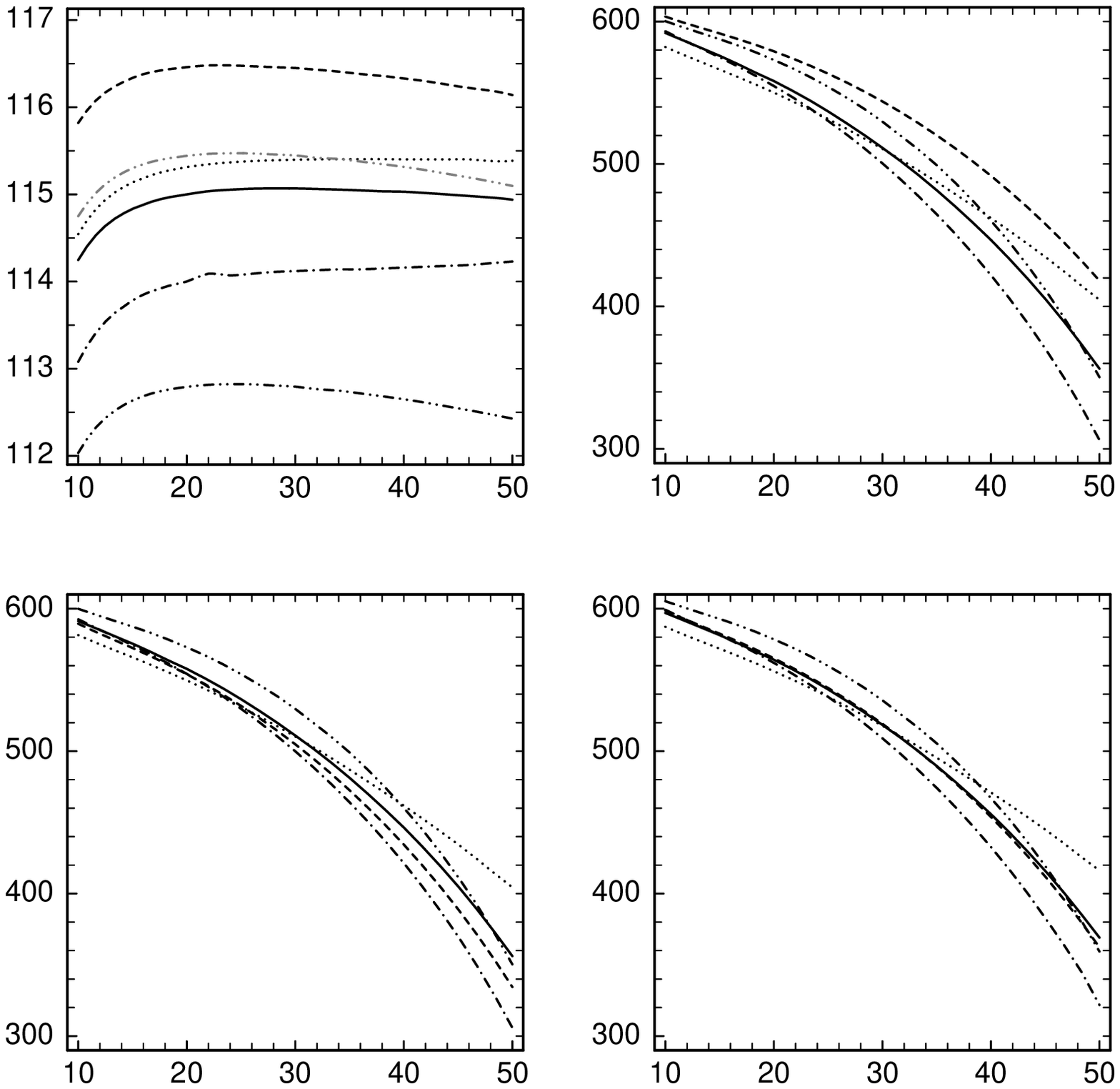, width=14cm}}
\put(1,100){\rotatebox{90}{$m$~[GeV]}}
\put(72,100){\rotatebox{90}{$m$~[GeV]}}
\put(1,30){\rotatebox{90}{$m$~[GeV]}}
\put(72,30){\rotatebox{90}{$m$~[GeV]}}
\put(38,1){\mbox{$\tan\b$}}
\put(108,1){\mbox{$\tan\b$}}
\put(38,71){\mbox{$\tan\b$}}
\put(108,71){\mbox{$\tan\b$}}
\put(61,128){\mbox{$h^0$}}
\put(128,126){\mbox{$H^0$}}
\put(60,56){\mbox{$A^0$}}
\put(128,56){\mbox{$H^\pm$}}
\end{picture}
\label{fig:higgsmasses}
\caption{Higgs boson masses as a function of $\tan\b$, for 
  $m_0=400$~Gev, $m_{1/2}=300$~GeV, $A_0=0$, $\mu>0$, $M_t=175$~GeV; 
  full (dotted) lines: \ISAJET~(7.58), dashed: \SOFTSUSY, 
  dash-dotted: \SPHENO, dash-dot-dotted: \SUSPECT\ (for 
  $h^0$, the grey dash-dot-dotted line corresponds to \SUSPECT\ + 
  FeynHiggsFast).}
}

\Fig{higgsmasses} shows the Higgs boson masses obtained by the four 
programs as a function of $\tan\b$ 
(the results obtained by \ISASUSYnn\,7.58 are again shown as 
dotted lines in \fig{higgsmasses}). 
Let us first discuss the masses of $A^0$ and $H^\pm$.
For $\tan\b=10$\,--\,50, we find differences in $m_A$ and $m_{H^\pm}$ 
of about 10\,--\,50~GeV, dominated over most of the $\tan\b$ range 
by the difference between \SPHENO\ and \SUSPECT. 
This has to be compared with differences of 100~GeV and more 
encountered with earlier versions, see for instance the 
dotted lines representing \ISAJETnn\,7.58. 
Assuming the error to be Gaussian, we now have 
$\Delta m_{A,H^\pm} \simeq \pm 10$~GeV at $\tan\b=25$ and 
$\Delta m_{A,H^\pm} \simeq \pm 20$~GeV at $\tan\b=50$. 
The bottom Yukawa coupling is, however, not the only source of 
differences in $m_A$. Another source is, for example, whether 
one uses running or pole values for masses in the calculation of one-loop
tadpoles $t_{1,2}$. Also, the scale and scheme of parameters in the one-loop
expressions for the tadpoles all vary.
These differences are formally of higher order and indeed each program has a
different approach.

The situation is somewhat different for the $(h^0\!,\,H^0)$ system 
because here additional radiative corrections are necessary. 
It is well known that these involve a theoretical uncertainty on 
$m_{h^0}$ of about 3~GeV~\cite{Degrassi:2002fi}, evidence of which 
can be seen in \fig{higgsmasses}. 
For completeness we note that \SUSPECT\ offers various choices of 
Higgs mass calculations. In \fig{higgsmasses}, we have used its 
default $m_{h^0}$ routine, i.e.\ {\tt ichoice(10)\,=\,0}, shown as 
a black dash-dot-dotted line. If we use instead \SUSPECT\ with 
FeynHiggsFast, {\tt ichoice(10)\,=\,3}, we get $m_{h^0}\sim 115$~GeV, 
shown as a grey dash-dot-dotted line in \fig{higgsmasses}. 
This will be relevant later in this paper when we discuss mSUGRA fits 
to LHC data.

\subsection{Focus point}

For large $m_0$, the running of $m_{H_2}^2$ becomes very steep and 
very sensitive to the top Yukawa coupling:
\begin{equation}
  \frac{dm_{H_2}^2}{dt} \sim \frac{3}{8\pi^2}\,h_t\,X_t +\ldots\,,\quad  
  X_t = (m_{\ti Q}^2 + m_{\ti U}^2 + m_{H_2}^2 + A_t^2)\,.
\end{equation}
As a result, the $\mu$ parameter given by 
\begin{equation}
   \mu^2 = \frac{\overline{m}_{H_1}-\overline{m}_{H_2}^2\tan^2\b}
                {\tan^2\b-1} - \frac{1}{2}\,M_Z^2  
\label{eq:muex}
\end{equation}
becomes extremely sensitive to $h_t$. 
This is visualised in \fig{htmu} where we show $h_t$ and $\mu$ as 
functions of $m_0$. The other parameters are $m_{1/2}=300$~GeV, 
$A_0=0$, $\tan\b=10$ and $\mu>0$. 
There is reasonable agreement on $\mu$ up to $m_0\sim 1$~TeV, 
and the differences observed for $m_0\lsim 2$~TeV are 
phenomenologically not so important. For larger values of $m_0$  
we observe, however, large discrepancies between the four programs. 
These lead to completely different chargino/neutralino properties 
for very large $m_0$, and likewise to very different excluded regions, 
depending on which program is used.
The situation is, however, already much better than the one reported 
in \cite{susy02}, c.f. the dashed lines of {\small ISASUSY\,7.58} 
in \fig{htmu} (earlier versions of the other codes also
gave quite different results).


\FIGURE[t]{%
\unitlength=1mm
\begin{picture}(135,68)
\put(0,2){\mbox{\epsfig{figure=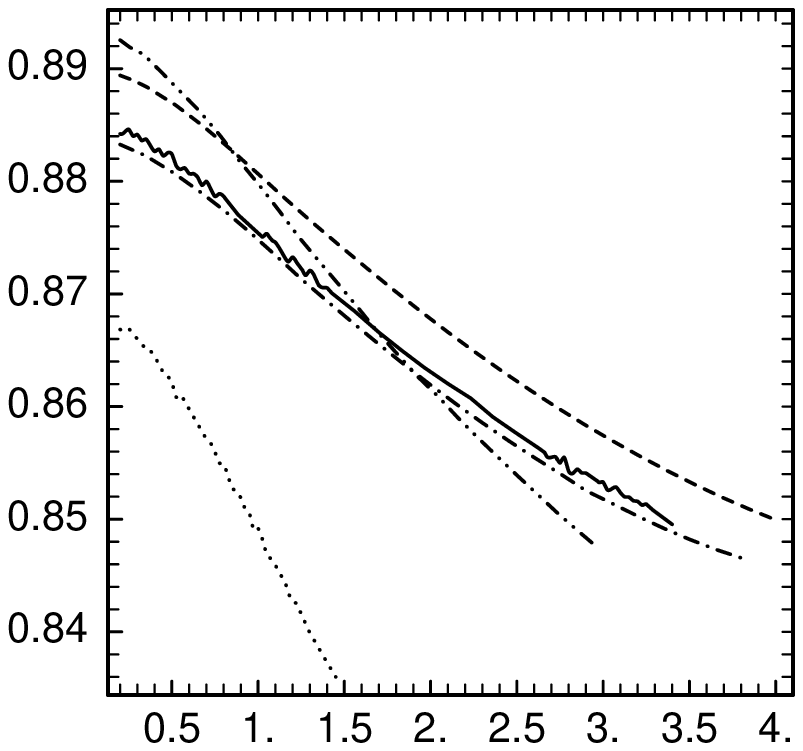,height=6.5cm}}}
\put(-6,29){\rotatebox{90}{$h_t(M_{SUSY})$}}
\put(28,0){\mbox{$m_0$~[TeV]}}
\put(74,2.5){\epsfig{file=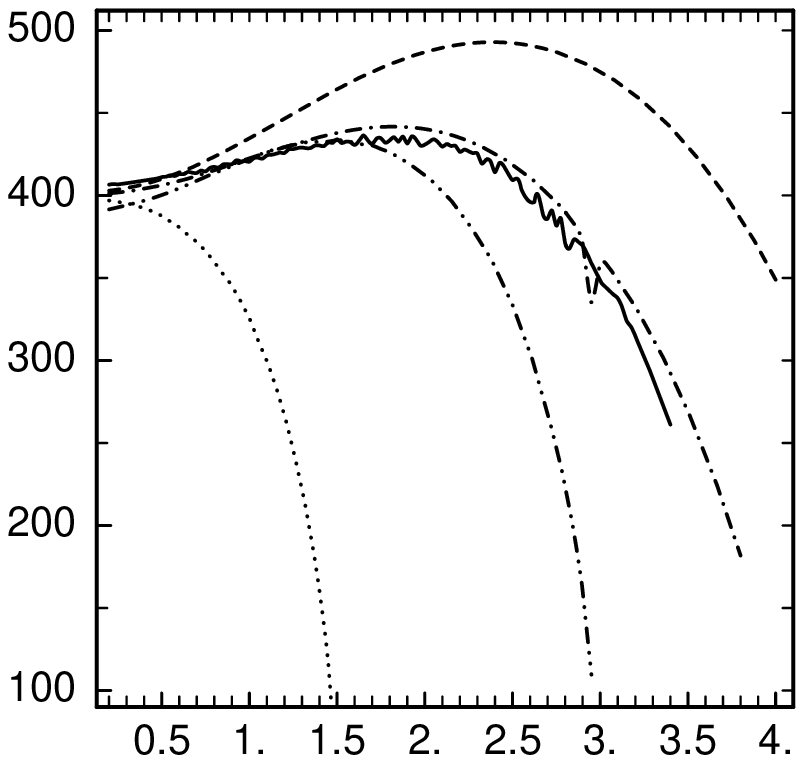, height=6.4cm}}
\put(69,31){\rotatebox{90}{$\mu$~[GeV]}}
\put(103,0){\mbox{$m_0$~[TeV]}}
\end{picture}
\label{fig:htmu}
\caption{$h_t$ and $\mu$ as a function of $m_0$ for $m_{1/2}=300$~GeV,
   $A_0=0$, $\tan\b=10$, $\mu>0$, $M_t=175$~GeV; 
  full (dotted) lines: \ISAJET~(7.58), dashed: \SOFTSUSY, 
  dash-dotted: \SPHENO, dash-dot-dotted: \SUSPECT.}}


\FIGURE[t]{%
\unitlength=1mm
\begin{picture}(132,68)
\put(0,2){\mbox{\epsfig{figure=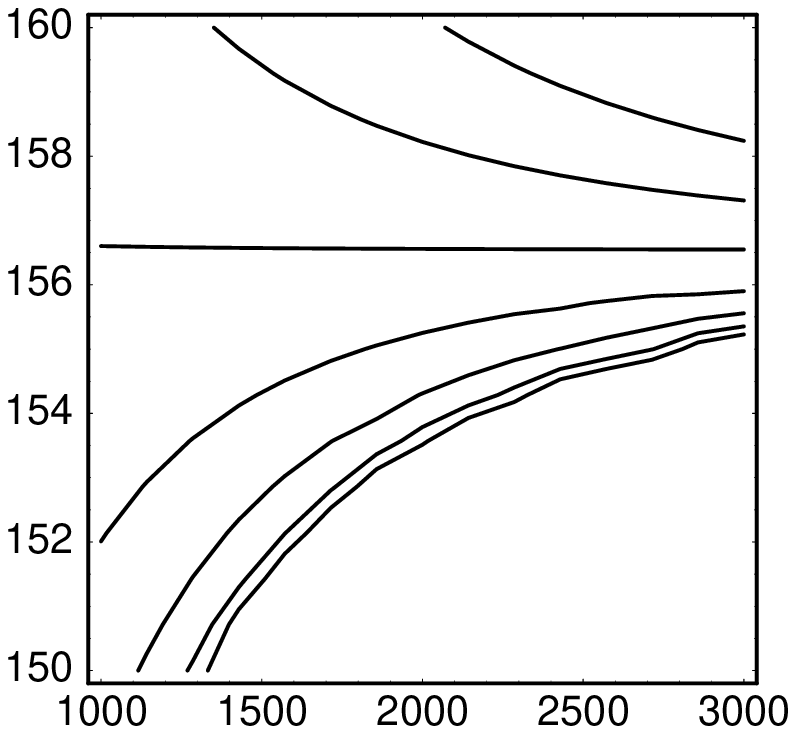,height=6.4cm}}}
\put(-6,28.5){\rotatebox{90}{$m_t$~[GeV]}}
\put(25,0){\mbox{$m_0$~[GeV]}}
\put(25.5,17){\mbox{\footnotesize 0}}
\put(16.5,25){\mbox{\footnotesize 200}}
\put(15,33){\mbox{\footnotesize 300}}
\put(15,46){\mbox{\footnotesize 400}}
\put(27,56){\mbox{\footnotesize 500}}
\put(48,57){\mbox{\footnotesize 600}}
\put(73,2){\mbox{\epsfig{figure=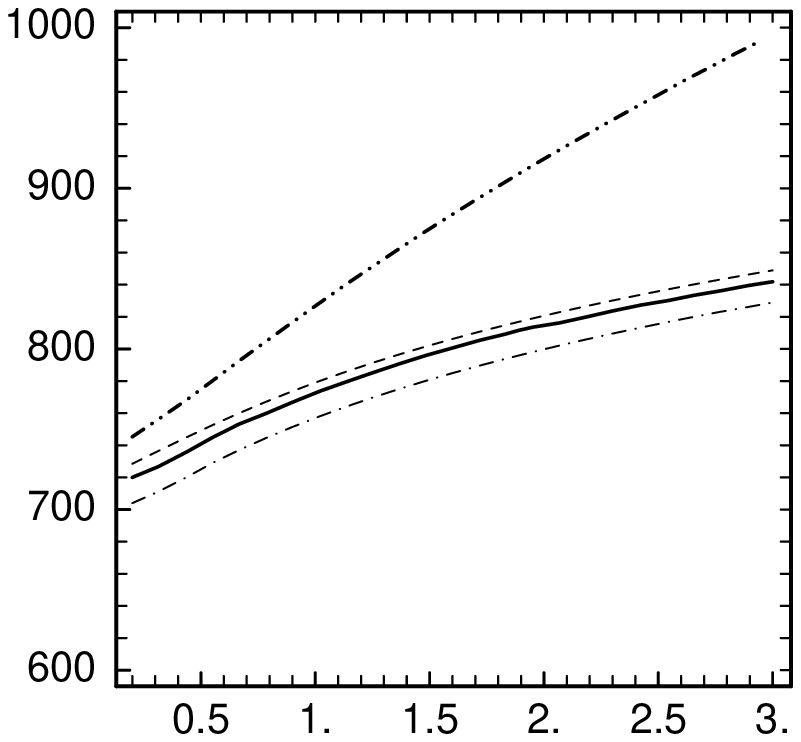,height=6.4cm}}}
\put(-6,60){(a)}
\put(67,60){(b)}
\put(67,29){\rotatebox{90}{$\msg$~[GeV]}}
\put(101,0){\mbox{$m_0$~[TeV]}}
\end{picture}
\label{fig:mum0mt}
\caption{(a) $|\mu|$ as given by eq.~\protect\eq{muap} for $m_{1/2}=300$~GeV,
   $A_0=0$, and $\tan\b=10$ in the 
  ($m_0,\hat m_t$) plane. 
  (b) Gluino mass as a function of $m_0$ for $m_{1/2}=300$~GeV,
   $A_0=0$, $\tan\b=10$, $\mu>0$, $M_t=175$~GeV; 
  full lines: \ISAJET\, dashed: \SOFTSUSY, 
  dash-dotted: \SPHENO, dash-dot-dotted: \SUSPECT.}}

In order to understand the behaviour of $\mu$ in \fig{htmu} 
it is useful to write eq.~\eq{muex} in the form 
\begin{equation}
   \mu^2 \simeq c_1\, m_0^2 + c_2\, m_{1/2}^2 - 0.5 M_Z^2\,.
\label{eq:muap}
\end{equation}
Approximate analytical expressions for $c_1$ and $c_2$ can be found 
e.g., in \cite{Carena:1994bv, Drees:1995hj}. 
For $A_0=0$ and $\tan\b=10$ we get \cite{Drees:1995hj}
\begin{equation}
   c_1 \sim\, \left(\frac{m_t}{156.5~{\rm GeV}}\right)^2-1\,,\qquad 
   c_2 \sim\, \left(\frac{m_t}{102.5~{\rm GeV}}\right)^2-0.52\,.
\label{eq:c1c2}
\end{equation}
\Fig{mum0mt}a shows contours of constant $\mu$ in this approximation 
in the $(m_0,\, m_t)$ plane. 
Notice the fast increasing dependence on $ m_t$ for increasing $m_0$. 
For $m_t\sim 156$\,--\,157~GeV, $\mu$ becomes almost 
independent of $m_0$. This is a signal of the actual focus point 
behaviour~\cite{Feng:1999mn}, which is defined as the insensitivity of
$m_{H_2}$ to its GUT scale value. Eq.~\ref{eq:muex} then implies that $\mu$ is
insensitive to $m_0$, since $m_{H_1}$ appears with a suppression factor of
$\tan^2 \beta-1$.
Interpreting $m_t$ in eq.~\eq{c1c2} as 
$m_t(M_{SUSY})=h_t(M_{SUSY})\,v_2(M_{SUSY})/\sqrt{2}$, we can directly 
relate the $m_0$ dependence of $\mu$ in \fig{htmu} to that of $h_t$.

Some more comments are in order. 
Firstly, \SUSPECT\ has only 1--loop RGEs for scalar SUSY parameters. 
For large $m_0$, the 2--loop terms lead to ${\cal O}(10\%)$ correction 
and should thus be taken into account. 
Secondly, 
as already mentioned in \sect{codes}, the sparticle masses that 
enter the radiative corrections have different values in different codes. 
In particular, there are large differences in the gluino masses 
between \SUSPECT\ and the other codes, as illustrated in \fig{mum0mt}b. 
Since the gluino mass enters $h_t$ via the $\sg\st$ correction eq.~\eq{dmt},  
this may account for the different slope of $h_t$ (and consequently of $\mu$)
in \SUSPECT\ compared to the other programs, as evident in \fig{htmu}.

It thus seems that for reliable results in the large $m_0$ region, 
a more complete calculation of the top Yukawa coupling  
at the 2--loop level is necessary.
This should include finite radiative corrections to all sparticle 
masses at the full 1--loop level.

\section{Comparison of theoretical and experimental uncertainties \label{sec:experiment}}

\subsection{Fits of mSUGRA parameters to LHC data \label{sec:lhc}}

In the ATLAS TDR \cite{ATLASTDR}, a case study was made of fitting 
the mSUGRA model 
to possible measurements of six reference scenarios. 
We have re-analysed these fits for two of these scenarios, LHC Point~1 
and Point~2. 
Here squarks and gluinos are produced with the dominant decays 
$\sg\to q\sq_{L,R}^{}$, $\sq_L^{}\to\nt_2 q\to\nt_1 hq$, 
$\sq_R^{}\to\nt_1 q$. 
The assumed measurements for the two points for low 
(${\mathcal L}=30$\,fb$^{-1}$) and high (${\mathcal L}=300$\,fb$^{-1}$)
luminosity are given in table~\ref{tab:edges}. They were estimated in
ref.~\cite{ATLASTDR} by using {\small ISAJET\,7.34} and simulating the 
ATLAS experiment to determine expected empirical errors.\footnote{It is 
beyond the scope of the present paper to re-perform the experimental 
analysis in order to have the numbers in table~\ref{tab:edges} more 
up-to-date.}

\TABULAR[b]{lllll}{
  \hline
    \bf Quantity \hspace*{1cm} & \qquad & \bf Low-L & \qquad & \bf High-L \\
  \hline
    $m_h$ (Point 1) && $\hphantom{0}95.4\pm 1.0$ GeV  && $\hphantom{0}95.4\pm 1.0$ GeV \\
    $m_h$ (Point 2) && $115.3\pm 1.0$ GeV && $115.3\pm 1.0$ GeV \\
    $m_{hq}^{\rm max}$  && $758.3\pm 25$ GeV  && $758.3\pm 25$ GeV \\
  \hline
    $m_{\sq_R}$     && $\hphantom{0}959\pm 40$ GeV    && $\hphantom{0}959\pm 15$ GeV \\
    $m_{\sg}$       && $1004\pm 25$ GeV   && $1004\pm 12$ GeV \\
    $m_{\st_1}$ (Point 1) && none && $647\pm 100$ GeV \\
    $m_{\st_1}$ (Point 2) && none && $713\pm 100$ GeV \\
  \hline}
{Possible LHC measurements for Point 1 and Point 2, from \cite{ATLASTDR}.
\label{tab:edges}}

With each of the programs under discussion  
we have performed a $\chi^2$ fit of the mSUGRA parameters $m_0$, 
$m_{1/2}$ and $\tan\b$ to the data of table~\ref{tab:edges}, 
taking $A_0=0$ and $\mu>0$. 
The results are listed in tables~\ref{tab:P1} and~\ref{tab:P2}. 
The quoted errors are at 1$\sigma$ (68.3\%~C.L.) from a simultaneous 
fit of all three parameters, i.e.\ $\Delta\chi^2=3.53$.
In case of \SUSPECT, we have used its default option, 
{\tt ichoice(10)\,=\,0}, for the calculation of the $h^0$ mass. 
When linking \SUSPECT~with FeynHiggsFast, {\tt ichoice(10)\,=\,3}, 
the results for $m_0$ and $m_{1/2}$ practically do not change. 
However, we get much lower values for $\tan\b$: 
$\tan\b=1.7\pm 0.1$ for Point~1, and 
$\tan\b=6.06\pm 2.06$ for Point~2 at high luminosity.

\TABULAR[t]{l|cccc}{
  \multicolumn{5}{c}{Point 1, Low-L} \\
  \hline
     & $\chi^2_{min}$ & $m_0$ [GeV] & $m_{1/2}$ [GeV] & $\tan\b$ \\
  \hline
    ISAJET   & 0.10 & $490\pm 135$ & $424\pm 25$ & $1.77\pm 0.21$ \\
    SOFTSUSY & 9.30 & $280\pm 246$ & $425\pm 32$ & $1.60\pm 0.03$ \\
    SPHENO   & 0.02 & $373\pm 175$ & $436\pm 26$ & $2.10\pm 0.15$ \\
    SUSPECT  & 0.32 & $411\pm 116$ & $410\pm 20$ & $2.08\pm 0.16$ \\
  \hline
\multicolumn{5}{c}{} \\
  \multicolumn{5}{c}{Point 1, High-L} \\
  \hline
    & $\chi^2_{min}$ & $m_0$ [GeV] & $m_{1/2}$ [GeV] & $\tan\b$ \\
  \hline
    ISAJET   & \hphantom{0}0.57 & $496\pm 61$ & $424\pm 12$ & $1.77\pm 0.20$ \\
    SOFTSUSY & 11.66            & $356\pm 78$ & $422\pm 12$ & $1.60\pm 0.03$ \\
    SPHENO   & \hphantom{0}0.27 & $370\pm 82$ & $436\pm 12$ & $2.10\pm 0.15$ \\
    SUSPECT  & \hphantom{0}1.79 & $422\pm 67$ & $409\pm 13$ & $2.08\pm 0.15$ \\
  \hline}
{Fit to LHC measurements for Point 1. \label{tab:P1}}

\TABULAR[t]{l|cccc}{
  \multicolumn{5}{c}{Point 2, Low-L} \\
  \hline
     & $\chi^2_{min}$ & $m_0$ [GeV] & $m_{1/2}$ [GeV] & $\tan\b$ \\
  \hline
    ISAJET   & 0.03 & $523\pm 129$ & $424\pm 24$ & $6.55\pm 2.37$ \\
    SOFTSUSY & 0.08 & $414\pm 140$ & $419\pm 23$ & $4.65\pm 0.76$ \\
    SPHENO   & 0.19 & $405\pm 167$ & $437\pm 26$ & $\gsim 7$ \\
    SUSPECT  & 0.06 & $444\pm 114$ & $409\pm 18$ & $\gsim 7$ \\
  \hline
  \multicolumn{5}{c}{} \\
  \multicolumn{5}{c}{Point 2, High-L} \\
  \hline
    & $\chi^2_{min}$ & $m_0$ [GeV] & $m_{1/2}$ [GeV] & $\tan\b$ \\
  \hline
    ISAJET   & 0.14 & $521\pm 58$ & $424\pm 11$ & $6.52\pm 2.30$ \\
    SOFTSUSY & 0.33 & $411\pm 68$ & $419\pm 11$ & $4.63\pm 0.88$ \\
    SPHENO   & 1.08 & $394\pm 80$ & $438\pm 13$ & $\gsim 7$ \\
    SUSPECT  & 0.20 & $450\pm 64$ & $408\pm 11$ & $\gsim 7$ \\
  \hline
}{Fit to LHC measurements for Point 2. \label{tab:P2}}

It is interesting to note that not only the central values but also 
the size of the errors can be quite different. Similarly, the minimum 
$\chi^2$, which is a measure of the quality of the fit, can show large 
variations.
To make the comparison easier and  to visualise correlations 
and non-Gaussian effects,  
we show in Figs.~\ref{fig:P1} and \ref{fig:P2} contours of 68\% and 
95\%~C.L. for the fits of Points~1 and 2 in the $(m_0,\,m_{1/2})$ 
and $(m_0,\,\tan\b)$ planes. The values of $\Delta \chi^2$ used for 
these confidence levels are based upon a simultaneous two-parameter fit, 
i.e. $\Delta \chi^2=2.3$ and 5.99.
The third parameter, $\tan\b$ or $m_{1/2}$, is always fixed to its 
best-fit value. 
As one can see, the error ellipses have only little or even no overlap.
We therefore conclude that the theoretical uncertainty is about the 
same size as the statistical one in the fitted quantities.

\FIGURE[t]{
\unitlength=1mm
\begin{picture}(132,66)
\put(0,2){\mbox{\epsfig{figure=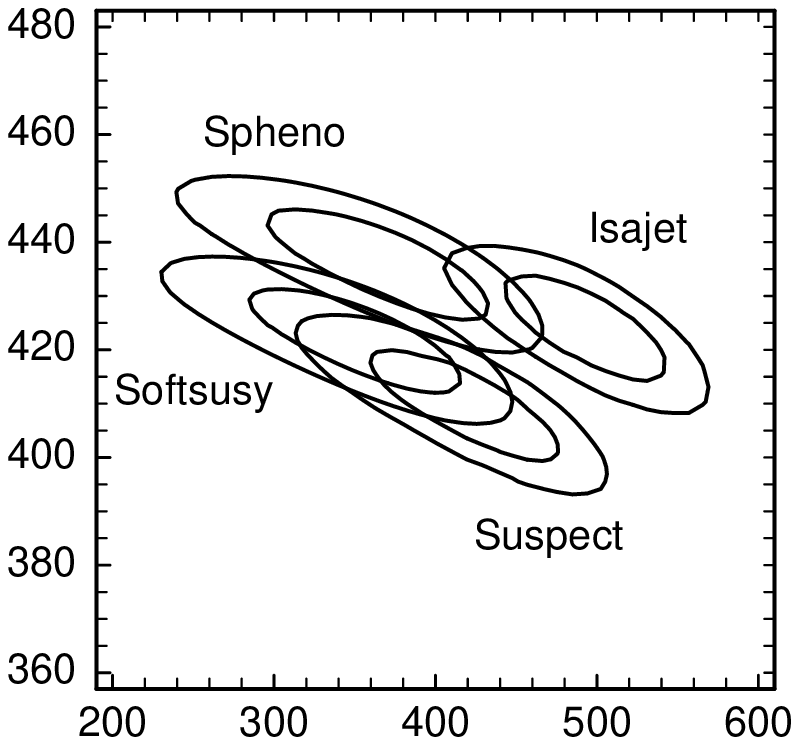,height=6.4cm}}}
\put(-6,28){\rotatebox{90}{$m_{1/2}$~[GeV]}}
\put(26,0){\mbox{$m_0$~[GeV]}}
\put(-7,60){(a)}
\put(75,2.5){\epsfig{file=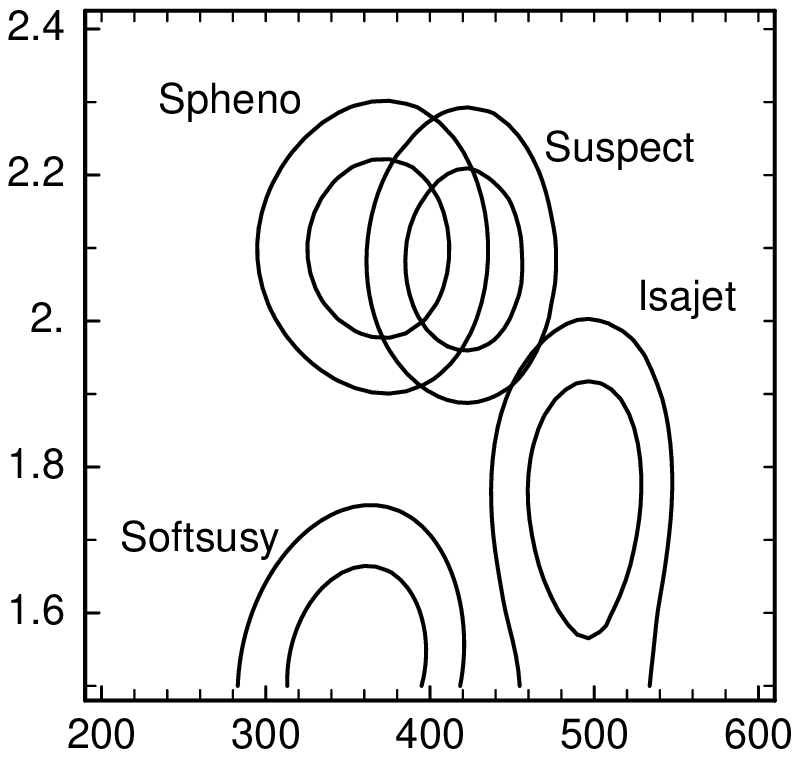, height=6.3cm}}
\put(70,32){\rotatebox{90}{$\tan\beta$}}
\put(100,0){\mbox{$m_0$~[GeV]}}
\put(68,60){(b)}
\end{picture}
\caption{Fit to LHC measurements for Point 1 with \ISAJET, \SOFTSUSY,
  \SPHENO~and \SUSPECT ~for ${\mathcal L}=300$~fb$^{-1}$. 
  Shown are contours of 68\% and 95\% C.L. for each program 
  in the $(m_0,\,m_{1/2})$ and $(m_0,\,\tan\b)$ planes,
  with the third parameter fixed to its best fit value, 
  c.f.~table~\ref{tab:P1}. 
\label{fig:P1} }}

\FIGURE[t]{
\unitlength=1mm
\begin{picture}(132,66)
\put(0,2){\mbox{\epsfig{figure=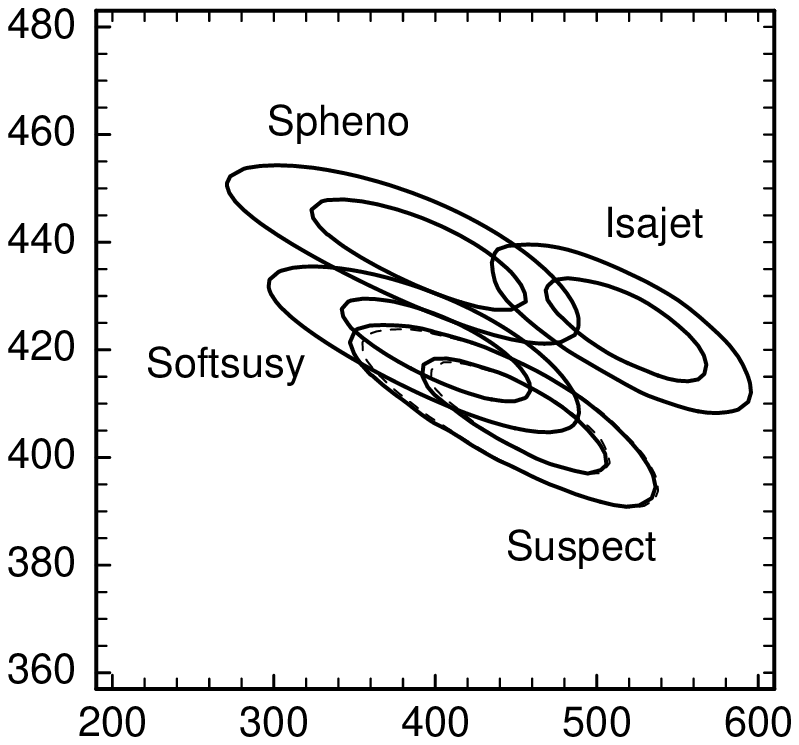,height=6.4cm}}}
\put(-6,28){\rotatebox{90}{$m_{1/2}$~[GeV]}}
\put(26,0){\mbox{$m_0$~[GeV]}}
\put(-7,60){(a)}
\put(75,2.5){\epsfig{file=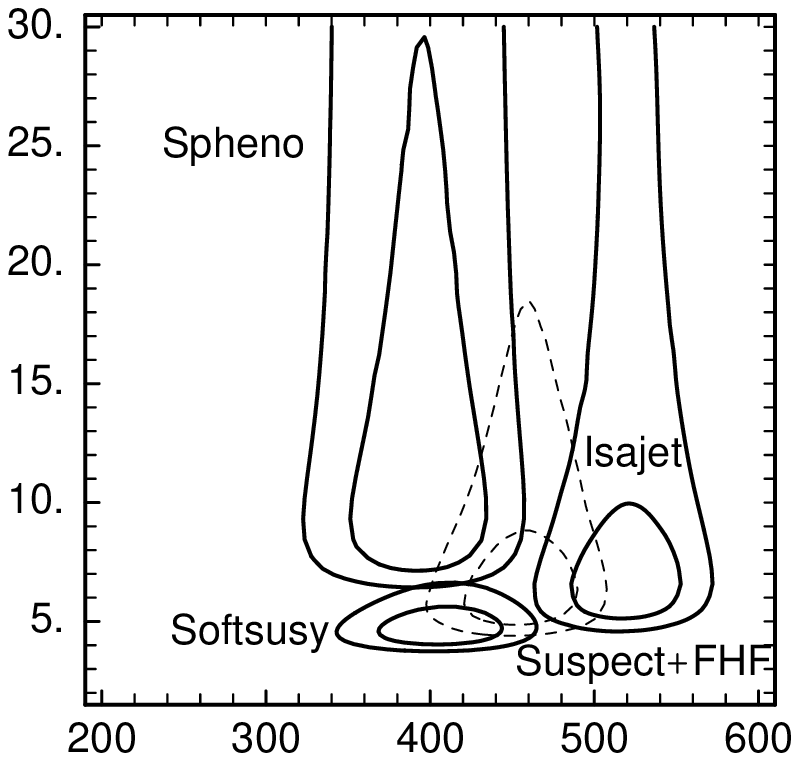, height=6.3cm}}
\put(70,32){\rotatebox{90}{$\tan\beta$}}
\put(100,0){\mbox{$m_0$~[GeV]}}
\put(68,60){(b)}
\end{picture}
\caption{Fit to LHC measurements for Point 2 with \ISAJET, \SOFTSUSY,
  \SPHENO~and \SUSPECT ~for ${\mathcal L}=300$~fb$^{-1}$. 
  Shown are contours of 68\% and 95\% C.L. for each program 
  in the $(m_0,\,m_{1/2})$ and $(m_0,\,\tan\b)$ planes,
  with the third parameter fixed to its best fit value, 
  c.f.~table~\ref{tab:P2}. 
  In case of \SUSPECT, the solid lines are for its default $m_h$ routine 
  and the dashed lines for $m_h$ calculated with FeynHiggsFast.
\label{fig:P2} }}

One might expect that the main source of these differences is the 
theoretical uncertainty on $m_h$. This is indeed the case for 
the determination of $\tan\b$. However, $m_h$ has only little influence 
on the fit of $m_0$ and $m_{1/2}$. This becomes clear when using {\it e.g.}, 
\SUSPECT~with different routines for the Higgs mass calculation. 
In \fig{P2}a, we show the results of \SUSPECT\,+\,FeynHiggsFast,  
{\tt ichoice(10)\,=\,3}, as dashed contours in addition to those obtained 
with its default  Higgs mass routine ({\tt ichoice(10)\,=\,0}, solid contours). 
In \fig{P2}b, we have omitted the default \SUSPECT~results 
to avoid confusion of the many lines. They would look similar to the 
\SPHENO\ contours but centred at $m_0=450$ GeV and without an upper limit 
on $\tan\beta$.
We thus conclude that the uncertainties in $m_0$ and $m_{1/2}$ 
mainly come from the differences in the programs pointed out in 
Sects.~\ref{sec:codes}--\ref{sec:tricky}, and not from  
different Higgs mass calculations.

We next address the question of how the theoretical uncertainty depends 
on the SUSY parameter point. 
Copious quantities of squarks and gluinos are expected to be produced 
at the LHC, leading to a fairly precise measurement of their masses, 
particularly if $m_{\chi_1^0}$ is determined accurately by a LC. 
In table~\ref{tab:mglu} we compare the $\sg$, $\ti u_L^{}$, $\ti u_R^{}$ 
and $\st_1$ masses obtained by the four programs for the Snowmass 
(SPS) points~\cite{Allanach:2002nj}. Assuming a Gaussian distribution, we
quote the variance  
of these masses as the theoretical error, i.e.\  
$\d m_X= \sqrt{\frac{1}{N-1} \sum_i\,[(m_X)_i-\overline{m}_X]^2}$ 
where $\overline{m}_X$ is the mean of $(m_X)_i$ and $N=4$ 
in our case. We make the following observations: 
(i) the absolute theoretical uncertainty (in GeV) varies from point to point;
(ii) the typical relative uncertainty in mSUGRA and mGMSB scenarios in
generic (i.e. not tricky) regions of parameter space
is about 2\,--\,5\%;
(iii) in some cases, in particular in focus point and mAMSB scenarios, 
the relative uncertainty is larger, about 5\,--\,10\%;
(iv) in any case, the theoretical error is of the same order of magnitude 
as the experimental one.

\begin{footnotesize}
\TABULAR[t]{c|ccccccccccc}{
  mass & code & 1a & 1b & 2 & 3 & 4 & 5 & 6 & 7 & 8 & 9 \\
  \hline 
  \boldmath $\sg$ 
  & ISAJET    & 607& 936& 794& 932& 732& 719& 718& 944& 835& 1296 \\
  & SOFTSUSY  & 614& 949& 802& 946& 743& 730& 729& 964& 852& 1306\\
  & SPHENO    & 594& 917& 782& 914& 719& 705& 704& 940& 836& 1232\\
  & SUSPECT   & 626& 964& 870& 959& 761& 730& 742& 986& 902& 1395\\
  & \bf error &\bf 13&\bf 20&\bf 40&\bf 19&\bf 18&\bf 12&\bf 16
              &\bf 22&\bf 31&\bf 67 \\
  \hline
  \boldmath $\ti u_L^{}$ 
  & ISAJET    & 536& 835& 1532& 817& 730& 642& 640& 858& 1079& 1233\\
  & SOFTSUSY  & 549& 851& 1582& 831& 753& 657& 662& 876& 1083& 1291\\
  & SPHENO    & 565& 876& 1563& 859& 764& 676& 674& 910& 1127& 1314\\
  & SUSPECT   & 570& 886& 1595& 867& 775& 681& 680& 910& 1138& 1502\\
  & \bf error &\bf 15&\bf 23&\bf 27&\bf 23&\bf 19&\bf 18&\bf 18
              &\bf 26&\bf 30&\bf 116 \\
  \hline
  \boldmath $\ti u_R^{}$ 
  & ISAJET    & 520& 807& 1529& 788& 714& 622& 626& 830& 1033& 1242\\
  & SOFTSUSY  & 569& 884& 1592& 866& 774& 681& 679& 914& 1142& 1297\\
  & SPHENO    & 548& 847& 1552& 828& 746& 655& 659& 880& 1080& 1266\\
  & SUSPECT   & 550& 852& 1585& 832& 754& 656& 662& 880& 1092& 1492\\
  & \bf error &\bf 20&\bf 32&\bf 29&\bf 32&\bf 25&\bf 24&\bf 22
              &\bf 35&\bf 45&\bf 114 \\
  \hline
  \boldmath $\st_1$ 
  & ISAJET    & 379& 633& 947& 621& 523& 236& 476& 774& 951& 998\\
  & SOFTSUSY  & 398& 658& 974& 645& 544& 232& 497& 813& 987& 951\\
  & SPHENO    & 398& 658& 964& 646& 545& 248& 497& 813& 982& 986\\
  & SUSPECT   & 410& 676& 1004& 663& 560& 243& 513& 831& 1015& 1140\\
  & \bf error &\bf 13&\bf 18&\bf 24&\bf 17&\bf 16&\bf 7&\bf 15
              &\bf 24&\bf 26&\bf 83\\
  \hline
}{Gluino and squark masses in GeV for the SPS benchmark points,   
  and their theoretical uncertainties.   
  The theoretical uncertainty is displayed in bold type face
  and is calculated as described in the text. 
\label{tab:mglu}}
\end{footnotesize}

\subsection{Linear Collider measurements \label{sec:lc}}
At a high-luminosity $e^+e^-$ Linear Collider, one expects to 
measure chargino, neutralino and slepton masses with accuracies at 
the per-cent or even per-mill 
level~\cite{Kuhlman:1996rc,Baer:1996vd,Freitas:2002gh,Aguilar-Saavedra:2001rg,Abe:2001gc}. 
We thus take the differences in these masses accessible at 
$\sqrt{s}=500~\gev$ as a measure of the theoretical uncertainty. 
Tables~\ref{tab:LC1} and~\ref{tab:LC2} compare masses obtained 
by the four programs for the various SPS points. 
Only those masses kinematically accessible at a 500 GeV LC are shown.
The error is again defined as the variance of the masses 
as in the previous section.

It turns out that the uncertainty in the LSP mass in mSUGRA and mGMSB 
scenarios is typically a few hundred MeV, depending on the parameter point. 
An exception is the focus point scenario (SPS2) where $\d\mnt{1}=1.4$~GeV.
For $\nt_2$ and $\ch_1$, we find uncertainties of about 1\,--\,3~GeV
in mSUGRA and mGMSB scenarios. 
(Here note that with earlier program versions, especially with 
ISASUSY\,7.51\,--\,7.63, we had discrepancies of 50\% and more  
for focus point scenarios.)
For the sleptons we find typical uncertainties of 1\,--\,2~GeV. 
We note that for SPS5,  the lighter stop would 
be accessible, with $m_{\ti t_1} = 235.6$, 232.3, 248.4 and 242.6 GeV 
for \ISAJET, \SOFTSUSY, \SPHENO~ and \SUSPECT, respectively, 
corresponding to an error of 7~GeV, c.f.~table~\ref{tab:mglu}.
In the mAMSB scenario, SPS\,9, we have much larger uncertainties 
of $\sim$\,8~GeV for $\mnt{1}$ and $\mch{1}$. This is due to the fact 
that \SOFTSUSY\ has 2-loop GUT-scale boundary conditions for mAMSB, 
while the other programs have only 1-loop boundary conditions. 
If we enforce 1-loop boundary conditions for all four programs the 
error decreases to 3~GeV. We therefore expect that if all programs were 
to use 2-loop boundary conditions the error would be better than 3~GeV.

By comparing table~\ref{tab:mglu} with tables~\ref{tab:LC1},\,\ref{tab:LC2} 
we see that the present theoretical uncertainty in mass predictions is 
significantly smaller for weakly interacting sparticles than for strongly 
interacting ones. This is expected, since the former have smaller threshold 
corrections. However, the errors in tables~\ref{tab:LC1},\,\ref{tab:LC2} 
are still larger by up to an order of magnitude than the expected 
experimental accuracies at an $e^+e^-$ Linear Collider.
Moreover, the differences in the masses produce cross sections and 
branching ratios that differ by a few per-cent. 
While this is not a problem for determining SUSY-breaking parameters 
at the low scale, it will be relevant when relating them to GUT-scale 
parameters in order to test their unification and to determine the 
sources of SUSY breaking.

\begin{footnotesize}
\TABULAR[t]{c|cccccccccc}{
point & code & $M_{\chi_1^0}$ & $M_{\chi_2^0}$ & $M_{\chi_1^\pm}$ & 
$M_{{\tilde \nu}_e}$ &  $M_{{\tilde \nu}_\tau}$ & $M_{{\tilde e}_R}$ &
$M_{{\tilde e}_L}$ &  $M_{{\tilde \tau}_1}$ & $M_{{\tilde
    \tau}_2}$ \\
\hline
1a & ISAJET & 96.6 & 176.8 & 176.4 & 185.8 & 184.9 & 142.9 & 201.9 & 
133.3 & 206.0\\ 
 & SOFTSUSY & 96.4 & 178.2 & 177.6 & 188.7 & 187.9 & 144.9 & 204.3 & 136.3 
& 207.8\\ 
 & SPHENO & 97.6 & 182.9 & 181.3 & 190.5 & 189.6 & 143.9 & 206.6 & 134.6 & 
210.3\\ 
 & SUSPECT & 97.4 & 179.8 & 179.1 & 188.5 & 187.5 & 144.9 & 204.4 & 135.7 
& 208.1\\ 
 & {\bf error} & {\bf 0.3} & {\bf 1.5} & {\bf 1.2} & {\bf 1.1} & {\bf 1.1} 
& {\bf 0.6} & {\bf 1.1} & {\bf 0.8} & {\bf 1.0}\\  \hline
1b & ISAJET & 161.2 & 299.2 & -- & -- & -- & -- & -- & 195.2 & --\\ 
 & SOFTSUSY & 161.3 & 303.5 & -- & -- & -- & -- & -- & 204.6 & --\\ 
 & SPHENO & 163.7 & 310.2 & -- & -- & -- & -- & -- & 195.1 & --\\ 
 & SUSPECT & 162.6 & 305.5 & -- & -- & -- & -- & -- & 199.7 & --\\ 
 & {\bf error} & {\bf 0.7} & {\bf 2.6} & {\bf --} & {\bf --} & {\bf --} & 
{\bf --} & {\bf --} & {\bf 2.6} & {\bf --}\\  \hline
2 & ISAJET & 120.2 & 221.7 & 221.5 & -- & -- & -- & -- & -- & --\\ 
 & SOFTSUSY & 118.6 & 232.0 & 231.8 & -- & -- & -- & -- & -- & --\\ 
 & SPHENO & 123.8 & 233.6 & 232.1 & -- & -- & -- & -- & -- & --\\ 
 & SUSPECT & 123.1 & 233.0 & 232.6 & -- & -- & -- & -- & -- & --\\ 
 & {\bf error} & {\bf 1.4} & {\bf 3.2} & {\bf 3.1} & {\bf --} & {\bf --} & 
{\bf --} & {\bf --} & {\bf --} & {\bf --}\\  \hline
3 & ISAJET & 160.5 & 297.0 & -- & -- & -- & 178.3 & 287.0 & 170.6 & 289.2\\ 
 & SOFTSUSY & 160.8 & 300.4 & -- & -- & -- & 182.0 & 290.4 & 175.3 & 
292.3\\
 & SPHENO & 162.7 & 307.2 & -- & -- & -- & 180.2 & 290.3 & 172.4 & 292.4\\ 
 & SUSPECT & 161.7 & 302.4 & -- & -- & -- & 182.1 & 290.7 & 174.8 & 
292.6\\ 
 & {\bf error} & {\bf 0.6} & {\bf 2.5} & {\bf --} & {\bf --} & {\bf --} & 
{\bf 1.0} & {\bf 1.7} & {\bf 1.3} & {\bf 1.6}\\  \hline
4 & ISAJET & 119.3 & 219.6 & 219.6 & -- & -- & -- & -- & -- & --\\ 
 & SOFTSUSY & 118.6 & 223.6 & 223.4 & -- & -- & -- & -- & -- & --\\ 
 & SPHENO & 121.4 & 228.9 & 227.4 & -- & -- & -- & -- & -- & --\\ 
 & SUSPECT & 121.1 & 225.9 & 225.7 & -- & -- & -- & -- & -- & --\\ 
 & {\bf error} & {\bf 0.8} & {\bf 2.3} & {\bf 1.9} & {\bf --} & {\bf --} & 
{\bf --} & {\bf --} & {\bf --} & {\bf --}\\  \hline
5 & ISAJET & 119.8 & 225.9 & 225.9 & 244.3 & 242.3 & 191.5 & 256.3 & 
180.9 & 257.7\\ 
 & SOFTSUSY & 118.7 & 229.4 & 229.2 & 248.2 & 246.1 & 193.5 & 259.4 & 
183.7 & 260.7\\ 
 & SPHENO & 121.1 & 230.0 & 229.8 & 248.0 & 245.9 & 192.0 & 259.4 & 181.7 
& 260.9\\ 
 & SUSPECT & 120.7 & 230.6 & 230.4 & 247.8 & 245.5 & 193.5 & 259.5 & 182.8 
& 260.6\\ 
 & {\bf error} & {\bf 0.6} & {\bf 1.2} & {\bf 1.2} & {\bf 1.1} & {\bf 1.0} 
& {\bf 0.6} & {\bf 0.9} & {\bf 0.7} & {\bf 0.9}\\  
\hline
}{Differences in predicted masses in GeV for the SPS points 1a to 5 
  (mSUGRA points). 
  Only sparticles with masses that kinematically can be produced 
  at a 500 GeV Linear Collider are displayed. 
  The theoretical uncertainty is displayed in bold type face
  and is calculated as described in the text. 
\label{tab:LC1}}
\end{footnotesize}

\begin{footnotesize}
\TABULAR[t]{c|cccccccccc}{
point & code & $M_{\chi_1^0}$ & $M_{\chi_2^0}$ & $M_{\chi_1^\pm}$ & 
$M_{{\tilde \nu}_e}$ &  $M_{{\tilde \nu}_\tau}$ & $M_{{\tilde e}_R}$ &
$M_{{\tilde e}_L}$ &  $M_{{\tilde \tau}_1}$ & $M_{{\tilde
    \tau}_2}$ \\
\hline
6 & ISAJET & 190.0 & 218.1 & 215.7 & -- & -- & 236.8 & -- & 227.8 & --\\ 
 & SOFTSUSY & 190.2 & 220.6 & 217.9 & -- & -- & 240.9 & -- & 232.8 & --\\ 
 & SPHENO & 191.2 & 225.3 & 222.3 & -- & -- & 237.6 & -- & 229.1 & --\\ 
 & SUSPECT & 190.2 & 222.2 & 219.6 & -- & -- & 240.9 & -- & 232.2 & --\\ 
 & {\bf error} & {\bf 0.3} & {\bf 1.7} & {\bf 1.6} & {\bf --} & {\bf --} & 
{\bf 1.3} & {\bf --} & {\bf 1.4} & {\bf --}\\  \hline
7 & ISAJET & 162.4 & 268.0 & -- & 248.7 & 248.3 & 127.3 & 261.1 & 119.9 & 
263.3\\ 
 & SOFTSUSY & 163.6 & 263.5 & -- & 247.2 & 246.9 & 126.4 & 259.3 & 120.5 & 
261.2\\ 
 & SPHENO & 163.4 & 271.1 & -- & 251.6 & 251.3 & 131.0 & 265.3 & 123.8 & 267.3\\ 
 & SUSPECT & 163.6 & 262.2 & -- & 246.9 & 246.6 & 127.8 & 259.4 & 121.6 & 
261.4\\ 
 & {\bf error} & {\bf 0.3} & {\bf 2.4} & {\bf --} & {\bf 1.2} & {\bf 1.2} 
& {\bf 1.2} & {\bf 1.6} & {\bf 1.0} & {\bf 1.6}\\  \hline
8 & ISAJET & 137.4 & 254.6 & -- & -- & -- & 175.7 & -- & 168.9 & --\\ 
 & SOFTSUSY & 138.4 & 261.2 & -- & -- & -- & 175.4 & -- & 169.9 & --\\ 
 & SPHENO & 139.2 & 266.1 & -- & -- & -- & 180.3 & -- & 173.5 & --\\ 
 & SUSPECT & 140.0 & 263.5 & -- & -- & -- & 177.6 & -- & 171.8 & --\\ 
 & {\bf error} & {\bf 0.6} & {\bf 2.8} & {\bf --} & {\bf --} & {\bf --} & 
{\bf 1.3} & {\bf --} & {\bf 1.2} & {\bf --}\\  \hline
9 & ISAJET & 174.8 & -- & 175.0 & -- & -- & -- & -- & -- & --\\ 
 & SOFTSUSY & 196.7 & -- & 196.7 & -- & -- & -- & -- & -- & --\\ 
 & SPHENO & 168.0 & -- & 168.4 & -- & -- & -- & -- & -- & --\\ 
 & SUSPECT & 167.3 & -- & 167.3 & -- & -- & -- & -- & -- & --\\ 
 & {\bf error} & {\bf 7.9} & {\bf --} & {\bf 7.9} & {\bf --} & {\bf --} & 
{\bf --} & {\bf --} & {\bf --} & {\bf --}\\  \hline
}{Differences in predicted masses in GeV for the SPS points 6 to 9 
  (6: non-minimal SUGRA, 7+8: mGMSB, 9: mAMSB). 
  Only sparticles with masses that kinematically can be produced 
  at a 500 GeV Linear Collider are displayed. 
  The theoretical uncertainty is displayed in bold type face
  and is calculated as described in the text. 
\label{tab:LC2}}
\end{footnotesize}

\section{Conclusions \label{sec:conc}}

If sparticles are detected at future colliders, measurements of their
properties (and confirmation that they are in fact sparticles) will take
place. The question following from this line of investigation will be: what
can we learn about SUSY breaking from these measurements? 
Is there a unification of certain SUSY breaking parameters, and is it 
possible to distinguish various SUSY breaking models? 

Vital ingredients for answers to these questions will be the amount 
and precision of empirical measurements made at the particular 
colliders~\cite{Kuhlman:1996rc,Freitas:2002gh,Aguilar-Saavedra:2001rg,Abe:2001gc,ATLASTDR,lesHouches}. 
Another vital ingredient will be the precision with which we can relate
the experimentally observed quantities at the TeV scale with the fundamental 
physics at the high-energy scale. Accurate extrapolations require multi-loop
results for the RGEs and the related threshold corrections.
In this paper, we have addressed the question: what is the current theoretical 
uncertainty associated with determining fundamental-scale SUSY parameters?
To this end, we compared four public state-of-the-art MSSM spectrum 
calculations: \ISAJET, \SOFTSUSY, \SPHENO\ and \SUSPECT,  
taking the spread of their results as a measure of the 
to-date uncertainty. Although this does not correspond to the usual notion of 
`theoretical uncertainty', it is pragmatic in the sense that (at least) one of
the available calculational tools will be used to perform fits if and when the
relevant data arrives.
The uncertainty was shown to be largest in certain tricky corners of parameter 
space: the focus point region and high $\tan \beta$. However, even in these 
regions, comparison with previous versions of the codes~\cite{susy02} shows 
that the theoretical uncertainty has significantly improved. 
Sparticle masses in these regions are particularly sensitive to the values 
of the Yukawa couplings (especially the top Yukawa for the focus point, 
and the bottom Yukawa for the high $\tan \beta$ regime). 
Slightly different treatments of top and bottom masses 
can lead to large differences in mass predictions.
It is therefore critical that the accuracy in deriving the running top 
and bottom $\DR$ Yukawa couplings is at a maximum in any calculation.

We used previous LHC estimates of expected empirical errors 
at two benchmark points to perform separate fits 
to mSUGRA with the four different codes. 
The parameters resulting from the four calculations show a difference 
comparable to the statistical error upon them, showing that theoretical 
uncertainties must be taken into account. 
We then went on to quantify the theory uncertainty associated with
squark and gluino masses for the SPS benchmark 
points.
The theory uncertainty was also quantified for parts of the
MSSM spectrum that would be kinematically accessible to a 500 GeV 
Linear Collider for the SPS points. 
The tables of spectra listed in the present article may be used 
for further comparisons with other (possibly new or private) computations.
The theoretical uncertainties for LHC and Linear Collider observables were
shown to significantly depend upon the  
point in parameter space being considered.
It will not be very precise, therefore, to use the theory uncertainties 
calculated here when
analysing actual data, since the SUSY breaking parameter point is 
{\it a priori} unknown. 

If sparticles {\it are} measured,
the correct modus operandi is clear when analysing data: 
any fits to a fundamental SUSY breaking model should be performed with several
state-of-the-art spectrum calculations in order to deduce the level of
theoretical uncertainty, which must then be included in final quoted errors. 
The theory errors must clearly be included when 
discussing the accuracy required to distinguish between SUSY models on
sparticle observables, such as in ref.~\cite{Allanach:2001qe}.

We note that each of the codes used assumed a pure MSSM desert up until the
GUT scale $\sim10^{16}$ GeV. 
We know already~\cite{Langacker:1995fk,Bagger:1995bw} that this
assumption  
is called into question because the gauge couplings themselves require
$\sim \mathcal{O}(1\%)$ corrections at the high scale in order to make them
unify properly.
It is thought that this level of correction is perfectly reasonable in, for
example, SUSY GUTs since model-dependent GUT-scale threshold 
corrections~\cite{Hisano:1992mh, Polonsky:1995rz} 
are expected from (for example) heavy coloured triplets. 
If the combined theoretical and empirical accuracy is significantly better
than $1\%$, then observables at colliders could be used to 
{\it measure} these 
threshold corrections.

Fortunately, theoretical errors in sparticle mass predictions
are certainly not static. 
There has been much progress reducing them recently (especially in the 
Higgs and electroweak symmetry breaking sectors, see for 
example~\cite{Brignole:2001jy,Degrassi:2002fi,Brignole:2002bz,
Martin:2002wn,Dedes:2002dy}). 
We expect this trend to continue, which, as our present results  
indicate, is desirable if we are to disentangle SUSY breaking from
experimental observables.

\acknowledgments

We thank Howie Baer, Abdelhak Djouadi, Fabiola Gianotti, Frederick James,  
Konstantin Matchev and Giacomo Polesello for useful discussions.
This work was partly funded by CNRS. 
W.P.\ is supported by the Swiss `Nationalfonds'.

\providecommand{\href}[2]{#2}\begingroup\raggedright\endgroup

\end{document}